\documentclass[fleqn,usenatbib]{mnras}

\usepackage{newtxtext,newtxmath}
\usepackage[T1]{fontenc}
\usepackage{tabularx}

\newcommand{\eazy}{{\tt{EAZY-py}}}
\newcommand{\bagpipes}{{\tt{Bagpipes}}}

\newcommand{\galfit}{{\tt{GALFIT}} }

\newcommand{\lya}{Ly$\alpha$ }

\newcommand\ha{H$\alpha$ }
\newcommand\hb{H$\beta$ }
\newcommand\sii{{\sc [Sii]}\ }
\newcommand\oiii{{\sc [Oiii]}\ }
\newcommand\nii{{\sc [Nii]}\ }
\newcommand\ciii{{\sc Ciii]}\ }

\newcommand\NeV{[\ion{Ne}{v}]\ }

\newcommand\HeI{\ion{He}{i}\ }

\newcommand\mkms{\rm km \, s^{-1} }
\newcommand\kms{$\mkms$ }

\newcommand\mflux{\rm \; erg \, s^{-1} \, cm^{-2} }

\newcommand\sfive{$\Sigma_5$ }

\DeclareRobustCommand{\VAN}[3]{#2}
\let\VANthebibliography\thebibliography
\def\thebibliography{\DeclareRobustCommand{\VAN}[3]{##3}\VANthebibliography}

\usepackage{graphicx}	% Including figure files
\usepackage{amsmath}	% Advanced maths commands

\title[Dual AGN at $z=5.4$]{A dual AGN at $z=5.4$ associated with a Lyman-alpha Nebula in the Center of a Cosmic Filament}

\author[Li et al.]{
Qiong Li\thanks{ qiong.li@manchester.ac.uk}$^{1}$,
Christopher J. Conselice$^{1}$, 
Qiao Duan$^{1}$,
Duncan Austin$^{1}$,
Tom Harvey$^{1}$,
Nathan Adams$^{1}$,
\newauthor
George Bendo$^{1}$,
Lewi Westcott$^{1}$,
Vadim Rusakov$^{1}$,
Zheng Cai$^{2}$,
Yuanhang Ning$^{3}$,
Shiwu Zhang$^{2}$
\\
$^{1}$ Jodrell Bank Centre for Astrophysics, University of Manchester, Oxford Road, Manchester M13 9PL, UK \\
$^{2}$ Department of Astronomy, Tsinghua University, Beijing, 100084, China \\
$^{3}$ Department of Scientific Research, Beijing Planetarium, Beijing, 100044, China
}

\date{Accepted XXX. Received YYY; in original form ZZZ}

\pubyear{\the\year{}}

\begin{document}
\label{firstpage}
\pagerange{\pageref{firstpage}--\pageref{lastpage}}
\maketitle

\begin{abstract}
Predictions from current theories and simulations suggest that dual AGN systems are exceedingly rare at high redshifts. The intense radiation and powerful outflows from AGNs regulate star formation, heat the interstellar medium, and drive massive gas outflows that shape the host galaxy and its surroundings. One manifestation of AGN feedback is the creation of extended Ly$\alpha$ nebulae. However, identifying these systems at high-$z$ is challenging. Here, we report a remarkable dual AGN candidate at $z \sim 5.4$ using JWST NIRCam and NIRSpec, with a separation of $\sim1.7$ arcseconds ($\sim10.4$ pkpc). This is one of the highest spectroscopically confirmed redshift dual AGNs discovered. Photometric SED fitting shows excellent agreement with AGN templates, strongly suggesting a rare dual AGN system. BPT diagrams and high ionisation lines further support the presence of AGNs. VLT/MUSE observations reveal strong extended Ly$\alpha$ emission, extending to $>22$ kpc, making it one of the most extended Ly$\alpha$ nebulae at $z \sim 6$. This provides observational evidence of anisotropic AGN-driven photoionization or shocks. The high Ly$\alpha$ escape fraction also indicates an AGN outflow. This dual AGN candidate is also associated with a well-defined overdensity, potentially at the center of a $z \sim 5.4$ protocluster or filamentary structure node. Further analysis indicates the fraction of dual AGNs is significantly higher than theoretically expected at high redshifts. This discovery provides a new opportunity to study dual AGN interactions and their impact on the circumgalactic medium and cosmic structure evolution.
\end{abstract}


%%%%%%%%%%%%%%%%%%%%%%%%%%%%%%%%%%%%%%%%%%%%%%%%%%

%%%%%%%%%%%%%%%%% BODY OF PAPER %%%%%%%%%%%%%%%%%%

\section{Introduction}

%Cosmological models of structure formation predict that supermassive black holes (SMBHs) in the close galaxy pairs, which eventually merge, should be widespread \citep{Puerto2025}. %These binaries, appearing as dual active galactic nuclei (AGNs) at kiloparsec separations, are expected in the universe. 
%At the early stages of their merger evolution, these systems can be identified as dual active galactic nuclei (AGNs), where the two AGN remain separated on kiloparsec scales and are not yet gravitationally bound.
Detecting dual active galactic nuclei (DAGNs) are identified as two active galactic nuclei (AGNs) in the early stages of a merger, remaining separated on kiloparsec scales and not yet gravitationally bound \citep{Perna2023a,Ishikawa2024}.  These are important systems for understanding a host of astrophysics including supermassive black holes (SMBHs) assembly across cosmic times \citep{Puerto2025,Junyao2024}. The extragalactic environments of these systems as well as their roles in gas accretion and SMBH mergers also provide insights into the connection between galaxy mergers, galaxy formation, AGN fueling, and local environment. 
%Their existence is predicted by cosmological simulations of structure formation, which suggest that galaxy mergers are more frequent in the early universe due to the higher galaxy density and smaller cosmic volumes at high redshift \citep{Hopkins2006}.

However, observational evidence for DAGN systems at high redshifts remains sparse, and very few cases are known.   Despite the expectation, only a handful of dual AGN systems have been identified at $z>1$, typically via gravitational lensing or serendipitous discoveries \citep{Mannucci2022, Silverman2020}.
The launch of JWST has however significantly advanced our ability to search for these high-redshift binaries. %We report the discovery of a unique dual AGN in this paper.
%Its NIRSpec and NIRCam instrument, with the exceptional sensitivity and sub-arcsecond resolution across the infrared spectrum, has opened new possibilities. 
Some recent research finds that high-redshift AGN frequently have close companions; for instance, \citet{Perna2023b} identified one AGN-galaxy pair candidates at $z=3.3$, while \citet{Maiolino2023BL} report three other pairs at $z>4$ that exhibit double-peaked H$\alpha$ emission providing evidence of harboring AGN.

Furthermore, simulations indicate that the number density of dual AGN separated by $<$30 pkpc should have a very low number density of $<10^{-5}$ comoving Mpc$^3$ at $z>5$\citep{Puerto2025}. At the same time, the lack of systematic surveys implies that high-$z$ dual AGNs remain exceedingly rare and difficult to identify.  However, because these dual AGN can reveal how star formation and AGN activity are triggered and fueled, they remain a missing part of our understanding of galaxy and black hole formation.

%30 photometrically confirmed dual AGN candidates have been identified in the Cosmic Evolution Survey (COSMOS-Web) spanning redshifts from 0 to 5, as presented in Li et al. (2024)

Driven by these inquiries, we have conducted a systematic search for galaxy pairs at $4.5 < z < 11.5$ using data from eight JWST NIRCam fields (total $\sim$189 arcmin$^2$, CEERS\citep{bagley2023ceers},   JADES\citep{eisenstein2023overview,bunker2023jades}, NEP\citep{Windhorst2022}, NGDEEP\citep{Bagley2023}, GLASS\citep{Treu2022}, El-Gordo\citep{Windhorst2022}, SMACS-0723\citep{Pontoppidan2022}, MACS-0416\citep{Windhorst2022}). Our serendipitous discovery of one dual AGN at $z\sim5.4$, the highest redshift spectroscopy confirmed dual AGN found to date, allows us to greatly increase our current understanding of galaxy mergers and the role of AGNs in pairs at high redshift for fueling black hole growth. This dual AGN represents the most exceptional case among all the mergers we have identified. In this system, both sources in the close pair have evidence of AGN activity and the pair are associated with an extended Ly$\alpha$ nebula. The pair is located within a galaxy group on larger scales, offering a rare opportunity to investigate early AGN fueling and the correlation between AGN activity and large-scale environment, whose properties we describe in this paper. These unique properties make this pair an exceptional target for investigating early-universe galaxy mass assembly, mergers, and feedback processes.

In this paper, we present the identification, spectroscopic confirmation, and physical characterization of this dual AGN system. We describe the data and observations used in Section~\ref{sec: data}, our selection and SED fitting methodology in Sections~\ref{sec: results - sed}, and discuss the implications of our findings in Sections~\ref{sec: results}. Finally, we summarize our conclusions in Section~\ref{sec: conclusion}. Throughout this paper, we assume a flat cosmological model with $\Omega_{\Lambda} = 0.7, \Omega_{m} = 0.3$ and $H_0 = 70 \ \mkms \, \rm Mpc^{-1}$ which implies the physical scale is $\approx 6.1 \, \rm kpc$ per arcsec at $z\sim5.4$, the redshift of this system. For the clarity in this article, we name the eastern galaxy JADES NIRCam ID: 32751 as G1 and the westernmost galaxy JADES NIRCam ID: 32927 as G2 in the paper.

\section{Observations and Data reduction }\label{sec: data}

We use a probabilistic pair counting methodology that integrates full photometric redshift posteriors to accurately quantify galaxy pairs which we describe in \citep{Duan2024}. Following the identification of potential merger pairs, we fit their NIRCam photometry and generate extensively vetted pair/merger catalogs. We next match these robust selected pair catalogs with NIRSpec spectra to select AGNs among them. Among 3452 galaxies at $z > 4.5$, we identified a total of 14 AGN-galaxy pairs at the projected distance r$_p$<30 kpc, with at least one galaxy covered by NIRSpec observations confirming its redshift. 

We find in total 14 pairs, however, the most notable among these is a $z \sim 5.4$ dual AGN candidate. The two sources G1 and G2 are separated by 1.7'' (projected as $\sim$ 10.4~kpc) at $z\sim5.4$ (see Fig.~\ref{fig: image}). Their redshifts are $z_{\rm spec}=5.440\pm0.001$ and $z_{\rm spec}=5.442\pm0.001$, with a comoving distance along the sight line of 0.52 $h^{-1}$Mpc. 

They are covered in the JADES\citep{ferruit2022near,bunker2023jades} and FRESCO surveys\citep{Oesch2023}. The dual AGN also has VLT/MUSE spectra\citep{Herenz2019} and ALMA observation. The broadband images of this dual AGN are from the JADES survey (PI: Eisenstein, N. Lützgendorf, ID:1180, 1210). For a rigorous evaluation of our photometry, we also use HST ACS images from CANDELS that cover the same field \citep{Grogin2011,Koekemoer2011} (Fig.~\ref{fig: image} a, c). The NIRSpec spectrum of G1 we use originates from the first data release of JADES \citep{ferruit2022near,bunker2023jades} featuring both G395H and PRISM spectroscopy. 
The spectra for G2 was obtained from the FRESCO survey \citep{Oesch2023}, which obtained R=1600 spectra within the NIRCam grism and F444W filter. The MUSE-Wide GTO program on the VLT also covers these sources, providing coverage of Ly$\alpha$.

\subsection{JWST NIRCam and HST Imaging observations}
\label{sec:data - jwst} 
%$z=5.450\pm 0.008$
%$z=5.490\pm 0.002$
%5.44012
%5.4417
The merger pair consists of G1 (
%ID: 32751,
RA: 03:32:29.52, DEC: -27:47:47.79, at $z=5.440\pm0.001$\citep{bunker2023jades}) and G2 (
%ID: 32927,
RA: 03:32:29.39, DEC: -27:47:47.46, at $z=5.442\pm0.001$\citep{Helton2024}). The two sources are detected in NIRCam observations of the GOODS-S field from the Deep tier of the JWST Advanced Deep Extragalactic Survey (JADES, PI: Eisenstein, N. Lützgendorf, ID:1180, 1210)\citep{Rieke2023}. Within this region there is imaging data from 9 NIRCam filters: F090W, F115W, F150W, F200W, F277W, F335M, F356W, F410M, and F444W.

%We process the images with the JWST Calibration Pipeline (version 1.8.2, CRDS v1084) with our custom steps, as described in EPOCHS II \citep{Adams2023}. We have considered various issues, including flat-field inaccuracies, cosmic ray, dark, distortion, bad pixel masks. 

We use the reduced imaging and catalogues of the JADES DR1 field that were produced by the EPOCHS project \citep{Conselice2024,Adams2023} to discover this system. To summarise, the field is reduced using a modified version of the official JWST NIRCam pipeline with calibration pmap 1084 in addition to the use of custom wisp correction templates and background subtractions. This results in $5\sigma$ depths between 29.8-29.6 mag in an aperture size of 0.32 arcsec in diameter. The final image resolution is 0.03''/pixel. The magnitudes in the NIRCam F277W band of G1 and G2 are 28.5 mag and 28.4 AB mag within an aperture of 0.32'' in diameter. 

We further enhance the photometric dataset by including imaging from the HST CANDELS survey \citep{Grogin2011,Koekemoer2011}, which covers a broader area in GOODS-S. 
%Within the GOODS-S DEEP region, the Hubble Ultra Deep Field (HUDF) offers exceptionally deep observations \citep{2013ApJ...763L...7E, 2013ApJS..209....3K, 2010ApJ...709L.133B, 2013ApJS..209....6I}.
We make use of HST ACS images in F435W($5\sigma \rm \ depth\sim28.7$mag), F606W($5\sigma \rm \ depth\sim28.9$mag), F775W($5\sigma \rm \ depth\sim28.4$mag), F814W($5\sigma \rm \ depth\sim28.7$mag), and F850LP($5\sigma \rm \ depth\sim28.2$mag)\citep{Grogin2011}. These images, derived from the CANDELS program, provide blue photometry for the JADES sources. 
%, as well as WFC3 image in F105W, F125W, and F160W

\subsection{JWST PRISM spectroscopy observation}\label{sec:data - prism}

The spectrum of G1 was obtained from the first JADES NIRSpec data release (PI: Eisenstein, N. Lützgendorf, ID:1180, 1210) \citep{ferruit2022near, bunker2023jades}.%, spanning the time-frame September 2022 to October 2022, with a focus on the publicly released data in the GOODS-S field. 
The PRISM data of G1, which covers a wavelength range of \(0.6 \, \mu\text{m}\) to \(5.3 \, \mu\text{m}\), with a spectral resolution of \( R \approx 30 - 330\) \citep{ji2022reconstructing}. During these observations, three micro-shutters were activated for the target. An exposure procedure was implemented consisting of a three-point nodding sequence along the slit, with each nod lasting 8403 seconds, and the entire sequence repeated four times. The subsequent extraction of flux-calibrated spectra was carried out using a specialized pipelines developed by both the ESA NIRSpec Science Operations Team and the NIRSpec GTO Team \citep{Bushouse2023}. A more detailed examination of the JADES/HST-DEEP spectra and the criteria used for sample selection is provided by \cite{eisenstein2023overview}. % Here we use the spectra reduced by the JADES team, and the details of process are provided in \citep{2023arXiv230602467B}. 
We attempt to use spectral data from the disperser-filter, specifically G395M/F290LP ($R \sim 1000$) and G395H/F290LP ($R \sim 27000$), but find the noise level to be approximately $50\%$ of the emission line strength. Thus, we continue our analysis using CLEAR/PRISM data.

\begin{figure*}
    \centering
    \includegraphics[width=\linewidth]{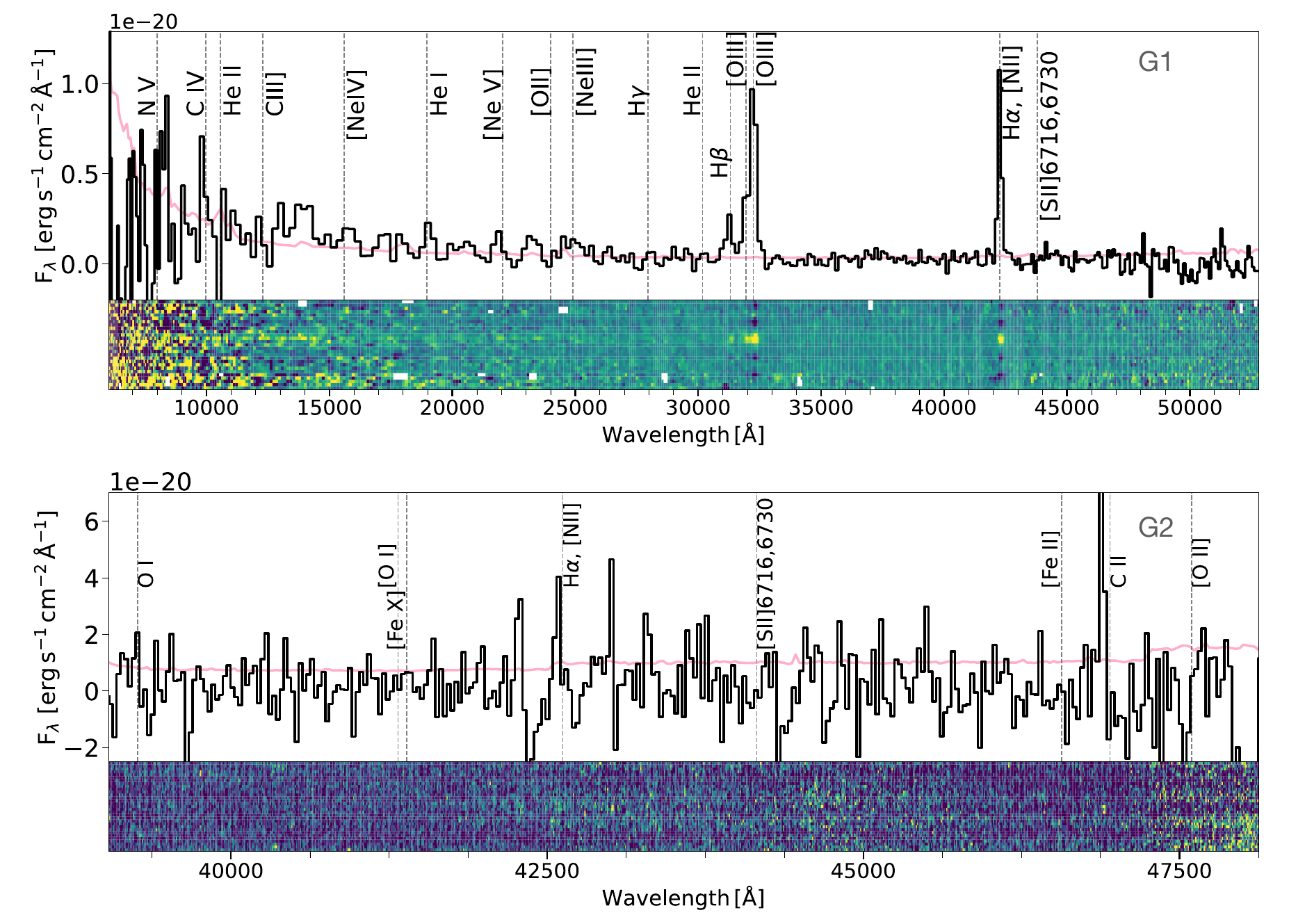}
    \caption{NIRSpec spectra of G1 (top) and G2 (bottom) taken with the NIRSpec $R\sim100$ PRISM. The associated uncertainty are marked in light red. The 2D spectra are shown below. Several prominent emission lines are noted. }
    \label{fig: spectrum two}
\end{figure*}

% fresco: https://arxiv.org/pdf/2304.02026.pdf
The spectrum of G2 is obtained from the JWST Cycle 1 program FRESCO (`First Reionization Epoch Spectroscopically Complete Observations', \citealt{Oesch2023}). FRESCO covers 62 arcmin$^2$ in each of the two GOODS/CANDELS fields, totaling 124 arcmin$^2$. Each field consists of $\sim2$ hr deep NIRCam/grism observations with the F444W filter, with $R \sim 1600$ and covering 3.8-5.0 $\mu$m. The 5$\sigma$ emission-line sensitivity reaches to $\sim2\times10^{-18} {\rm erg\,s^{-1} cm^{-2}}$. This is achieved with 8 grism exposures of 880s taken with the MEDIUM2 readout mode, with 9 groups for each pointing. This results in a total grism exposure time of 7043 s at each position. Four large scale dithers are used, but with no sub-pixel dithering. The reduced spectrum of the NIRCam/grism data are kindly provided by Jakob Helton and processed using the publicly available grizli code \footnote{ https://github.com/gbrammer/grizli}. The 12 pixel central gap is used to minimize self-subtraction \citep{Kashino2023}. Continuum subtracted spectra are created using a running median filter along each row. The spectrum are shown in Figure~\ref{fig: spectrum two}.

\begin{figure*}
\centering
\includegraphics[width=0.95\textwidth]{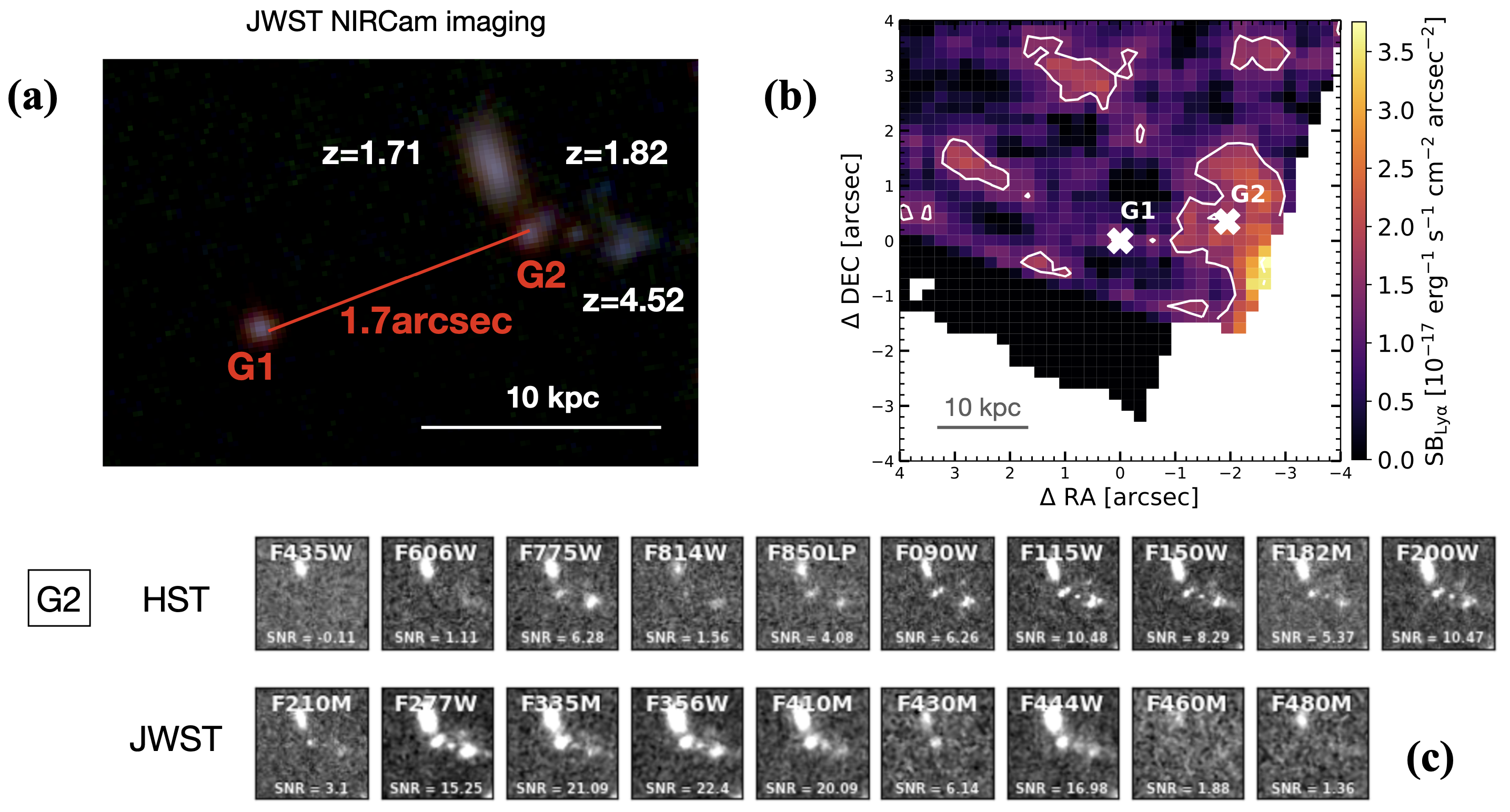}
\caption{
\textbf{(a)} JWST NIRCam RGB images (R: JWST/F444W; G: JWST/F277W; B: JWST/F115W) of the dual AGN we discovered. 
\textbf{(b)} G2 is surrounded by a Ly$\alpha$ nebula extending $\sim$22 pkpc, making it one of the most extended Ly$\alpha$ nebulae observed at $z\sim6$. This excited ionizaated gas suggests the presence of a possible AGN-driven outflow. This is the continuum subtracted pseudo-narrow band Ly$\alpha$ image from VLT/MUSE, spanning {$\sim \pm 2000$ km s$^{-1}$} with central wavelength $\lambda_{\rm NB} = \lambda_{\rm Ly\alpha} (1+z_{\rm spec}) \approx 7831 $\AA, and Gaussian smoothed spatially by $0.5''$. The color map and the contours indicate the Ly$\alpha$ flux density and the signal-to-noise ratio (+2, 3, 4$\sigma$), respectively.  The $2\sigma$ surface brightness limit is $9.1\times 10^{-18}$ erg s$^{-1}$ cm$^{-2}$ arcsec$^{-2}$. The Ly$\alpha$ emission surrounding G2 extends beyond the MUSE observation boundaries. Therefore, the measured extent represents a lower limit. \textbf{(c)} G2 exhibits clumpy morphologies on a very small scale ($< 0.1$'' in F090W band, $< 0.6$ pkpc, Fig.\ref{fig: image} c ). This indicates possible past merger activity inside the galaxy, aligning with the scenario in which merging triggers the AGN activity.
}
\label{fig: image}
\end{figure*}

\subsection{VLT MUSE IFU}\label{sec:data - muse}

The MUSE-Wide guaranteed time observations (GTOs) survey with the ESO-VLT covers these two sources. The IFU spectral range covers 4750\AA\, to 9350\AA\, which allows for the detection of \lya at 1215.67\AA\, in the redshift range of $2.9 < z < 6.7$. The seeing conditions for MUSE-Wide were around 1'' for most of the field (the full width at half maximum, FWHM, of the point spread function).  We use this data to trace the Lyman-$\alpha$ line across our system.   %The characteristic \lya luminosity for the luminosity function of log$L_{\star}[{\rm erg/s}] = 42.2^{+0.22}_{-0.16}$ \citep{Herenz2019}. 
The data and data products we used, such as cut-outs, mini-cubes and extracted spectra as well as emission line catalogues are publicly available\citep{Herenz2019}{\footnote{https://
musewide.aip.de/project/}}.

\subsection{ALMA}\label{sec:data - alma}

The ALMA Science Archive currently contains data covering these objects in two bands.  The ALMA Band 3 (84.2-115.0~GHz) data were acquired in ALMA programs 2017.1.00138.S and 2021.1.00547.S.  The ALMA Band 6 (244.1-275.0~GHz) data were acquired in ALMA programs 2015.1.00098.S, 2015.1.00543.S, and 2017.1.00755.S.  All of these observations consist of mosaics covering large sections of the GOODS/CANDELS fields. The data were processed using the {\sc Common Astronomy Software Applications} ({\sc CASA}) \citep{McMullin2007}.  

The raw visibility data were first calibrated either using the pipeline calibration restoration scripts (with different versions of {\sc CASA} used for different datasets for compatibility reasons) or, for older data acquired before the ALMA pipeline became operational, using manual calibration scripts in {\sc CASA} version 6.5.4.  Both types of scripts first apply amplitude corrections for the system temperature, phase corrections based on water vapor radiometer measurements, and antenna position corrections.  After this, the scripts then derive bandpass corrections to correct the phases and amplitudes as a function of frequency, then next derive phase and amplitude corrections as a function of time, and then apply those corrections.

The images were created using {\sc tclean} in {\sc CASA} version 6.5.4.  For the imaging, we only used fields (i.e., pointings) that covered the two targets in this paper, and only small subregions of these fields were imaged.  The centre of each image was set to the centre of G1.  To optimize the signal-to-noise in the images for detecting faint sources, we used natural weighting and the Hogbom deconvolver \citep{Hogbom1974}.  The pixel scales were set so that the full width and half-maximum (FWHM) of each beam was sampled by $>5$ pixels.  Continuum images were created using data from all observed frequencies in each ALMA band.  Image cubes covering the CO (5-4) and CO (6-5) lines were created using just the spectral windows that covered the line emission.  First, the continuum was subtracted from the visibility data while avoiding channels that could potentially include line emission from either object.  The continuum-subtracted data were then used to create image cubes with the rest frequencies set to the positions of the lines corresponding to the expected line centres at $z=5.44$.  The image cube channels have a width of 100 km s$^{-1}$ and extend to $\pm$2500 km s$^{-1}$ from the centre of the spectral line or the edge of the spectral window containing the line.  Details about the continuum images are given in Table~\ref{tab:almacont}, and information about the spectral line image cubes is provided in Table~\ref{tab:almaline}.

\begin{table*}
\centering
\small
\caption{ALMA continuum image details.}
\label{tab:almacont}
\begin{tabular*}{\linewidth}
{@{\extracolsep{\fill}}lcccccc}
\hline
Band &
  Frequency &
  Central &
  Beam &
  Pixel &
%  \multicolumn{2}{c}{Image Size} &
  Image Size &
  RMS %&
%  Calibration 
\\
&
  Coverage &
  Frequency &
  FWHM &
  Scale &
%  (pixels) &
  (arcsec) &
  Noise %&
%  Uncertainty$^a$ 
\\
&
  (GHz) &
  (GHz) &
  (arcsec) &
  (arcsec) &
%  & 
  &
  (mJy/beam) %&
  \\
\hline
3 &
  84.2 - 115.0 &
  99.6 &
  $1.22 \times 0.92$ &
  0.1 &
%  $500 \times 500$&
  $50 \times 50$&
  0.015 %&
%  5 
  \\
6 &
  244.1 - 275.0 &
  295.6 &
  $0.21 \times 0.19$ &
  0.02 &
%  $1000 \times 1000$ &
  $20 \times 20$ &
  0.035 %&
%  10 
  \\
\hline
\end{tabular*}
%$^a$ This information is from the ALMA Technical Handbook \citep{Cortes2024}, which is available from \url{https://almascience.eso.org/documents-and-tools/cycle11/alma-technical-handbook}.
\end{table*}

\begin{table*}
\centering
\small
\caption{ALMA spectra line image cube details$^a$.}
\label{tab:almaline}
\begin{tabular*}{\linewidth}{@{\extracolsep{\fill}}lcccccc}
\hline
Band &
  Reference &
  %Number &
  \multicolumn{2}{c}{Channel} &
  Frequency &
  Velocity &
  RMS \\
&
  Frequency &
  %of &
  \multicolumn{2}{c}{Width} &
  Coverage &
  Coverage &
  Noise \\
&
  (GHz)$^b$ &
  %Channels &
  (GHz) &
  (km/s) &
  (GHz) &
  (km/s) &
  (mJy/beam/channel) \\
\hline
CO (5-4) &
  89.353 &
  %42 &
  0.0298 & 
  100 &
  88.608 - 89.830  &
  -1600 - 2500 &
  0.5 \\
CO (6-5) &
  107.216 &
  %33 &
  0.0358 & 
  100 &
  106.322 - 107.466  &
  -700 - 2500 &
  0.4 \\
\hline
\end{tabular*}
\begin{tabularx}{\linewidth}{@{}X@{}}
$^a$ The beam sizes, pixel scales, image sizes, and calibration uncertainties are either the same as or very similar to what is listed in Table~\ref{tab:almacont} for the Band 3 continuum data.\\
$^b$ This is the observed frequency of the line corresponding to $z=5.44934$.
\end{tabularx}
\end{table*}

%\section{Target Selection}
%We initially search for merger pairs with projection separations of less than 50 kpc, utilizing the best-fit photometric redshifts of each galaxy. For line-of-sight selections, we combine the full photometric redshift probability distributions of galaxies that meet the projection separation criteria. This allows us to calculate the probability $\mathcal{N}_{z}$ that a potential pair is a real pair, following the approach outlined in \citep{2015A&A...576A..53L, 2017MNRAS.470.3507M, 2019ApJ...876..110D}. Here, we describe the method for calculating $\mathcal{N}_z$. For detailed descriptions of each step and the complete high-redshift merger catalogs, please refer to \cite{2024arXiv240709472D}.

%For two galaxies with normalized redshift probability distributions $P_{1}(z)$ and $P_{2}(z)$, their combined redshift probability distribution is defined as 
%\begin{equation}
%\mathcal{Z}(z) = \frac{2 P_{1}(z) P_{2}(z)}{P_{1}(z) + P_{2}(z)}.
%\end{equation}
%$\mathcal{N}_z$ is the integral of $\mathcal{Z}(z)$ over the entire redshift range:
%\begin{equation}
%\mathcal{N}_z = \int_{0}^{\infty} \mathcal{Z}(z) \, dz.
%\end{equation}

%Our dual AGN merger candidates have an $\mathcal{N}_z$ value of 0.999 and projection separations of $10.4$ pkpc, indicating a very high probability of being a true merger pair. Subsequent NIRSpec spectra confirmed that their redshifts are indeed consistent, with a redshift difference of only $\Delta z = 0.008$.

\section{SED Fitting and Morphology Measurements }\label{sec: results - sed}

%We collected data from 10 broadband filters including HST and JWST (ACS/F435W, F606W, F814W, and NIRCam/F090W, F115W, F150W, F200W, F277W, F356W, F444W) and two medium-width filters (JWST NIRCam observations: F335M, F410M). Since both objects are point sources and to avoid contamination from nearby foreground galaxies, we utilized a consistent aperture size of 0.32'' in diameter for magnitude measurements.
% WFC3/F105W, F125W, F140W, F160W,

\subsection{\texorpdfstring{\eazy} {Lg} SED modeling and AGN templates}
\begin{figure*}
    \centering
    \includegraphics[width=\linewidth]{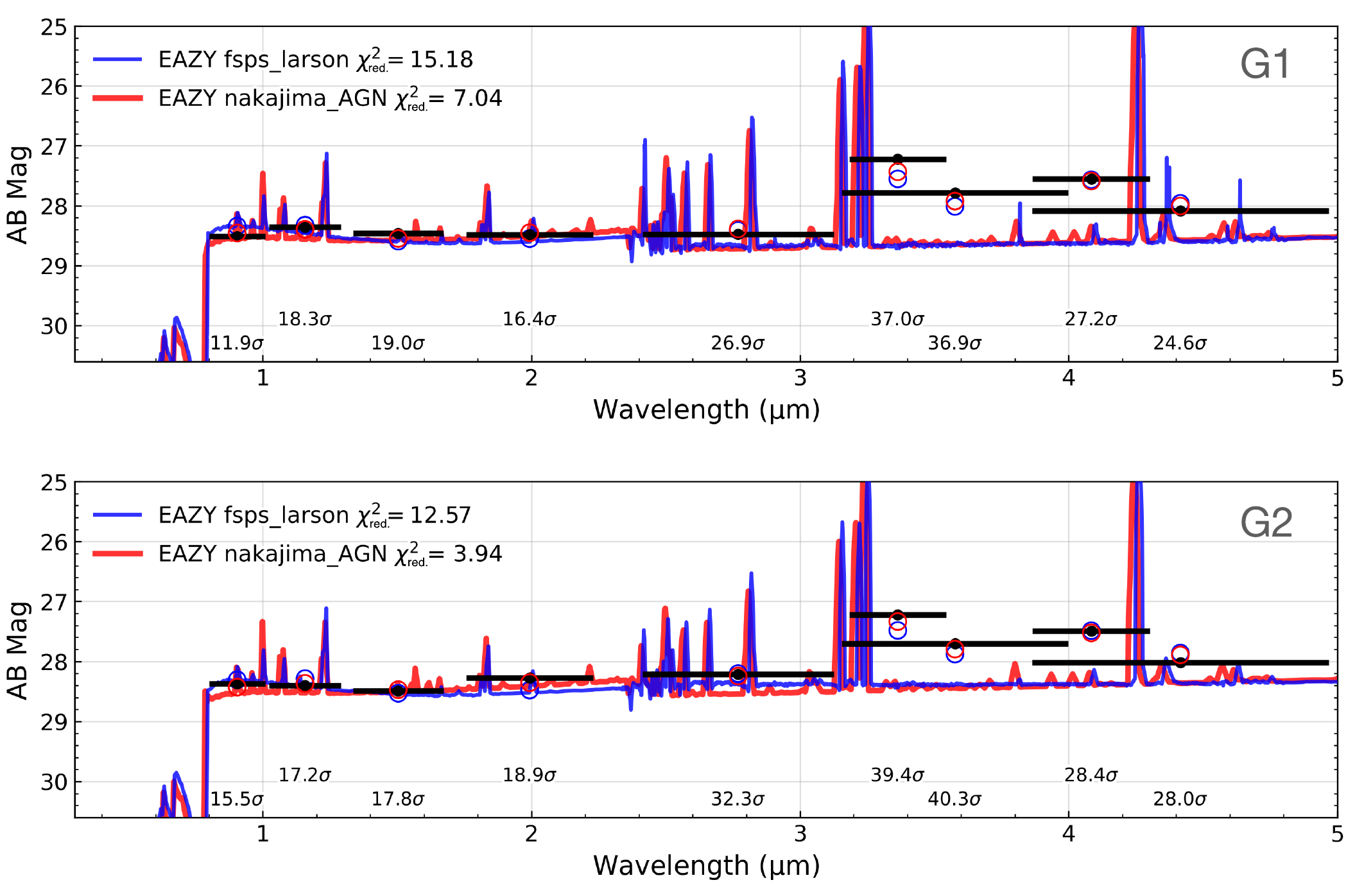}
    \caption{SED fitting of G1 and G2 using \eazy{} with two SED templates. The FSPS+Larson set \citep{larson2022spectral} is normal galaxy templates. The Nakajima set is AGN + star formation template, combined the direct collapse black hole (DCBH) hosts from \citep{Nakajima2022} and the FSPS+Larson set.}
    \label{fig: SED}
\end{figure*}

To perform SED modeling with the photometry data, we first use the \eazy{}, photometric redshift code \citep{brammer2008eazy,brammer2021gbrammer}. We use the same templates FSPS+Larson and parameters as in previous papers of the EPOCHS series (e.g. \citealt{Adams2023,Conselice2024}). The bluer templates we used are described in \citet{larson2022spectral}, and we utilize the tweak\_fsps\_qsf\_v12 template sets with a Chabrier initial mass function for our analyses, as detailed in \citep{Conroy2010,Bruzual2003} and \citep{Chabrier2003}, respectively. We apply the IGM attenuation derived from \citep{1995ApJ...441...18M}. We consider the dust effects using the prescription of \citep{calzetti2000dust}. We allow for $E(B-V)$ values up to 3.5 to account for potential very dusty galaxies at these high redshifts and to assess likely errors arising from low redshift contamination. 
The SED fitting is shown in Figure~\ref{fig: SED}.

To assess the presence of AGN components, we reprocessed our sources using \eazy{} with a combination of SED templates, including a template set for direct collapse black hole (DCBH) hosts from \citealt{Nakajima2022} and the FSPS+Larson set \citep{larson2022spectral}, following the method in \citep{Ignas2023}. These templates are specifically optimized for unobscured, intermediate-mass ($10^5 - 10^6M_{\odot}$) active black holes, expected to contribute significantly to high-redshift AGN. This AGN+star formation template set are referred to as the `Nakajima' set. We use the templates with UV power-law slopes ($\alpha$) and Big-Bump temperatures ($T_{\rm bb}$) at: $\alpha$ = -1.2, -1.6, -2.0, and $T_{\rm bb}$ = $5 \times 10^4, 1 \times 10^5, 2 \times 10^5$ K. We fix ionization parameter log U of -0.5,
%and metallicity $Z = 0.014$, 
as reasonable choices for high-redshift. The fitting results reveal that the reduced $\chi_{\rm red.}^2$ values with the Nakajima AGN templates for G1 and G2 are 7.04 and 3.94, respectively. Their AGN fractions ($f_{\rm agn}$) both approach 1. The $\chi_{\rm red}^2$ using the AGN template is notably lower by $\Delta\chi_{\rm red.}^2>8$ compared to the one obtained with the galaxy template, confirming that the enhancement in fit quality upon adding the AGN models. This strongly suggests that both sources harbor AGNs. Further evidence supporting the presence of AGNs will be presented through emission-line diagnostics in subsequent sections. 

\subsection{\texttt{Bagpipes} SED fitting with spectroscopy and photometry}
We employ \texttt{Bagpipes} \citep{carnall2018inferring, carnall2019vandels} with a simple AGN model to simultaneously fit photometric and spectroscopic galaxy data. The AGN model has a broken power law and broad \ha and \hb components. We set a log-uniform prior on equivalent width of $1000$\AA$\leq\rm EW_{\rm H\alpha \ \& \ H\beta} < 5000$\AA.

Different star formation histories (SFHs) can significantly impact the inferred stellar mass and SFR for the measured systems \citep{furtak2021robustly, 2022ApJ...927..170T, harvey2024epochs, wang2024quantifying}. We present results from our fiducial parametric SFH model - the log-normal SFH - and one non-parametric SFH, the continuity model \cite{leja2019measure}. These models are widely accepted and have been utilized in various works \citep[e.g.,][]{2023Natur.619..716C, 2023MNRAS.524.2312E, 2023MNRAS.522.6236T, 2023MNRAS.519.5859W, 2023arXiv230602470L}. The redshift is fixed to the spectroscopic redshift.

We implement a two-component dust law from \cite{charlot2000simple}, as implemented in \cite{carnall2018inferring}, with separate dust attenuation components for young stars ($\leq 10$ Myr) and older stellar populations, with an absorption curve slope (A$_{\rm V}\propto \lambda^-n$) of n$=0.7$, allowed to vary with a Gaussian prior with a standard deviation of 0.3. We use uniform priors for dust and metallicity. We set prior limits for metallicity in the range of $[\text{0}, 3.0] \, \text{Z}_{\odot}$, dust prior in the range of $[0., 6.0]$ mag in $A_\text{V}$, ionization parameter $\mathrm{log}_{10}U$ in range of $[-3, -1]$. The timescale for star formation is constrained to stop at $t_\text{U}$, with $t_\text{U}$ denoting the age of the Universe at the spectroscopic redshift. In addition, \cite{kroupa2001MNRAS.322..231K} IMF, \cite{2003MNRAS.344.1000B} SPS model, and the  \cite{calzetti2000dust} dust attenuation model is implemented. Different SFH timescales can significantly impact the inferred SFRs \citep{wang2024quantifying, 2024MNRAS.tmp..884D, harvey2024epochs}, so we present SFRs averaged over both 10 Myr and 100 Myr timescales.

To simultaneously fit spectroscopy and photometry we additionally fit a 3rd order polynomial to scale the observed spectra to the photometry, in order to account for slit losses and calibration offsets, although this component is typically constrained in the fit as a small linear offset without strong wavelength dependence. We also fit an additional noise component on the spectra, with a log uniform prior between (1, 10) to account for underestimated uncertainties. 

\subsection{Bursty Star Formation}
\begin{figure}
    \centering
    \includegraphics[width=0.9\linewidth]{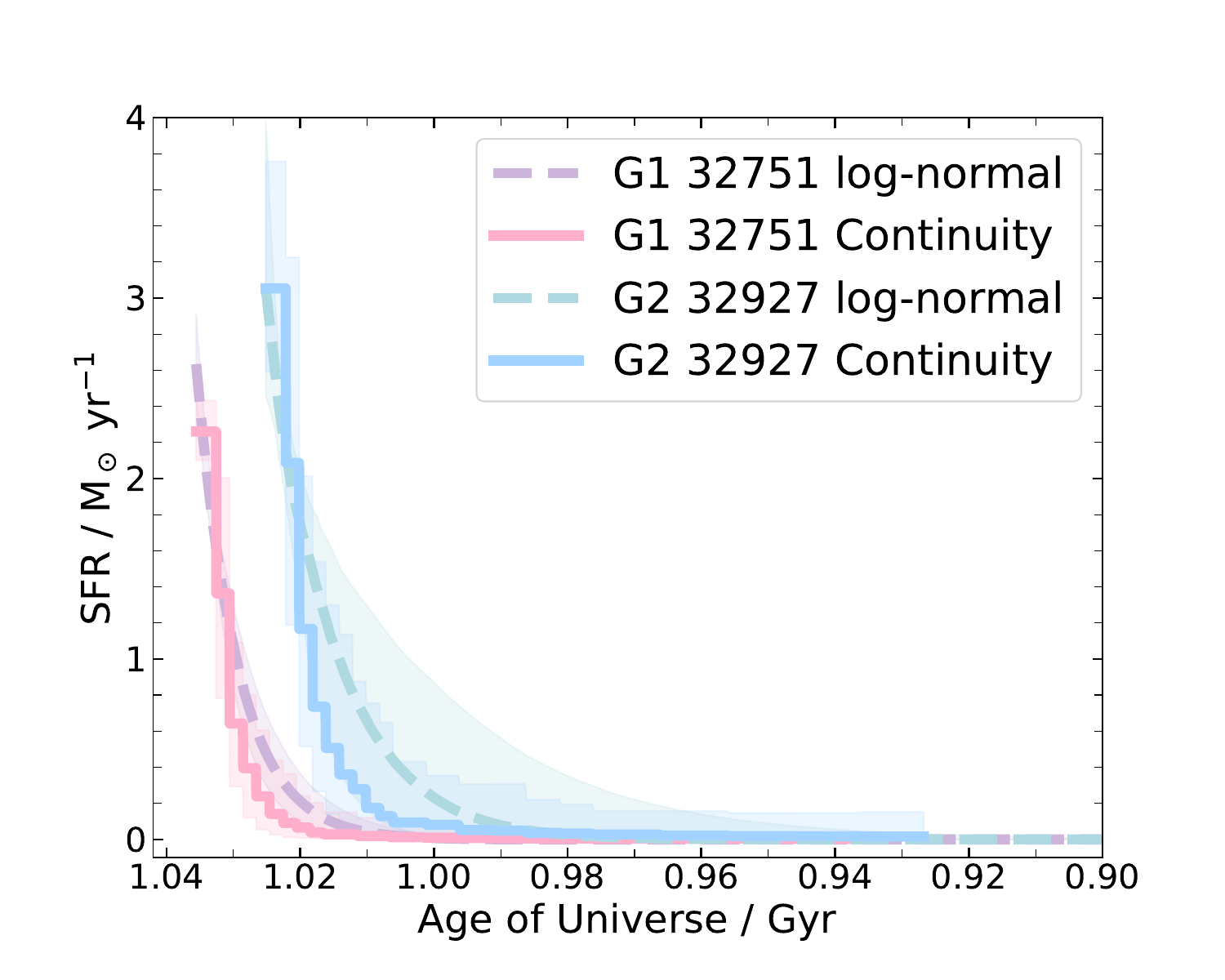}
    \caption{The SFH of G1 and G2 constructed from two different SFH models. The dashed lines are log-normal SFH, and solid lines are Continuity SFH. The galaxies G1 and G2 involved in the merger are currently both in the state of a starburst.}

    \label{fig: SFH}
\end{figure}

We find that the star-forming properties of both systems exhibit a rising or bursty nature, shown in Figure~\ref{fig: SFH}, as indicated by the log-normal SFH results including an AGN in the fitting. The SFRs averaged over 100 Myrs are quite small at  $0.19_{-0.03}^{+0.04}$ and $0.33_{-0.11}^{+0.25}$ for AGN 1 and AGN 2, respectively, and $1.52_{-0.16}^{+0.15}$ and $2.04_{-0.25}^{+0.23}$ when using a 10 Myr average. Although a higher SFR from the 10 Myr average aligns with expectations from the log-normal SFH model, which suggests an increasing SFR as galaxies evolve, a factor of approximately seven times higher than $\mathrm{SFR}_{\mathrm{100Myr}}$ strongly indicates a relatively recent starburst event. We then employ our second SFH model, a non-parametric Continuity model \citep{2019ApJ...876....3L}, to validate this finding. We do not include any bursty components in the Continuity SFH when running the fits, yet the SFH reconstructed from this model also displays a bursty nature. The SFH of the two galaxies constructed from two SFH models are shown in \autoref{fig: SFH}. The dashed lines are log-normal SFH, and solid lines are Continuity SFH. For both galaxies, the SFH constructed from the Continuity model is in great agreement with the log-normal SFH. The resulting specific star formation rates log(sSFRs) are $\sim -7.07^{+0.03}_{-0.04}$ and $\sim -7.15^{+0.12}_{-0.24}$ consistent with other low-mass, high-redshift galaxies, further supporting a recent bursty star formation episode.

\subsection{UV Morphology}
We use \galfit version 3.0.5 \citep{Galfit1,Galfit2} to fit a single Sersic light profile to each galaxy. \galfit is a least-squares-fitting algorithm that finds the best-fitting model by minimizing the reduced chi-squared statistic ($\chi_{\rm red.}^2$). We choose filters that closely match the rest-frame optical wavelength of the sources to minimize the morphological $k$-correction effect \citep{TaylorMager2007}. The Sérsic profile we use adopts the following form 
\begin{equation}
    I(R)=I_e \exp \left\{-b_n\left[\left(\frac{R}{R_e}\right)^{1 / n}-1\right]\right\},
    \label{eqn:Sérsic}
\end{equation}
where $I(R)$ is the intensity at a radial distance $R$ from the galaxy's center. $R_{e}$ is the galaxy's half-light radius, marking the radius where 50\% of the total luminosity is encompassed. $I_e$ is the intensity at this half-light radius. The Sérsic index $n$ controls the shape of the galaxy's light profile \citep{Sersic,ciotti_1991,caon_1993}.  Additionally, $b_{n}$, derived from $n$, can be computed according to \citep{ciotti1999}. The derived $R_{e}$ of G1 is $0.31\pm0.03$kpc 
%(i.e. $\sim0.05''$)
, the Sérsic index $n$ for this system is $0.51\pm0.47$. We fixed the Sérsic index $n=1$ to achieve a better fit for G2. $R_{e}$ of G2 is $0.72\pm0.02$kpc (i.e. $\sim0.12''$). The NIRCam F444W point-spread function (PSF) has an FWHM of approximately 0.14'', suggesting that both G1 and G2 exhibit compact UV continuum morphologies.

Notably, G2 exhibits clumpy morphologies on a very small scale ($< 0.1$'' in F090W band, $< 6.1$ pkpc, Fig.\ref{fig: image} c). This may indicate past merger activity or a burst of star formation inside the galaxy, aligning with the scenario in which a merger is triggering the AGN activity \citep{Hopkins2006}. There are no further sources around G1. G2 is surrounded by three foreground sources, with $z_{\rm phot}= 4.52, 1.82$ and 1.71 respectively (Fig.\ref{fig: image} a).

\begin{figure*}
    \centering
    \includegraphics[width=\linewidth]{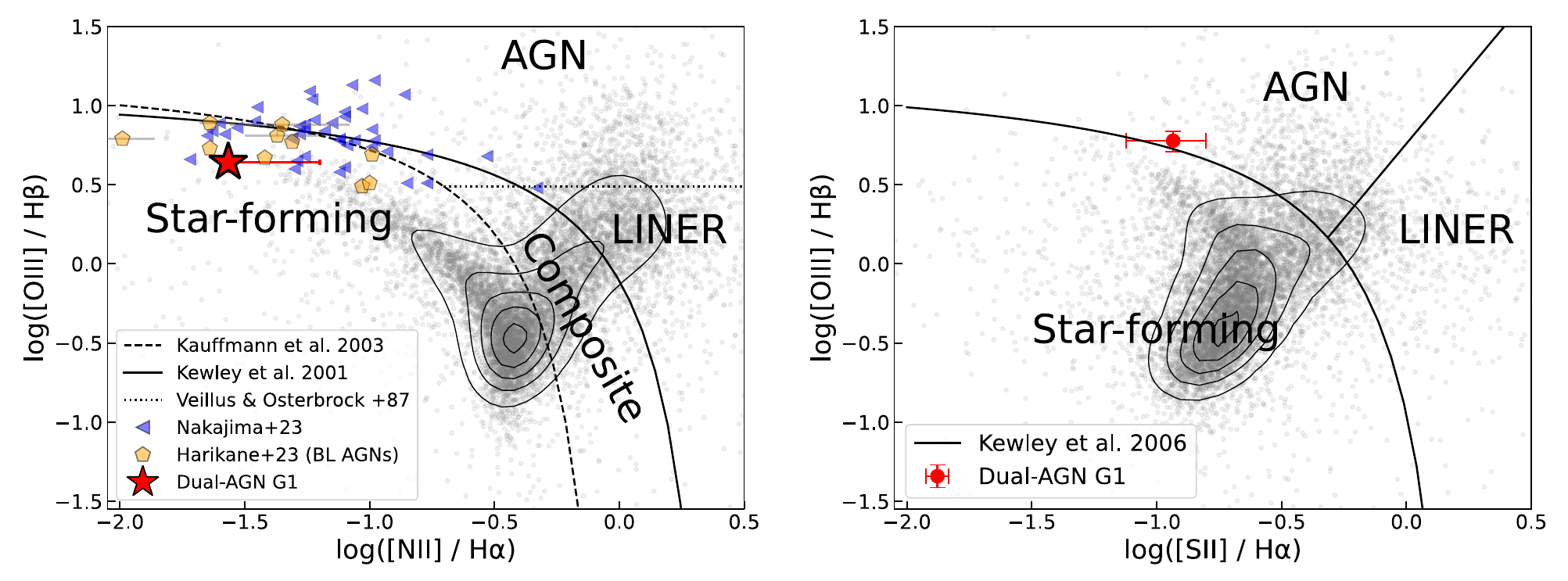}
    \caption{The BPT emission-line diagnostic diagram. {\it Left:} The dashed and solid lines denote the $z = 0$ and $z = 2.3$ boundary between the star-forming and AGN regions of the diagram defined by \citet{Kauffmann2003} and \citet{Kewley2001}, respectively. {\it Right:} The line denotes the boundary between star-forming and AGN regions as defined in \citet{Kewley2006}. The contours and background gray points are from 927,552 low-z SDSS DR7 galaxies \citep{2009ApJS..182..543A}.}
    \label{fig: BPT}
\end{figure*}

\section{Results}\label{sec: results}

\subsection{BPT emission-line diagnostic diagram and high-ionization lines}

We compare the line ratios of \nii$\lambda$6584/\ha and \sii$\lambda$6730/\ha with \oiii$\lambda$5007/\hb, classifying the galaxy based on its position in each plot. 
The grism spectrum of G2 from FRESCO survey is in the F444W band, which covers only \ha; whereas the JADES prism spectrum of G1 covers \ha+\nii and \hb, thus analysis for this diagnostic is focused solely on G1.
For G1, limited by the resolution of NIRSpec PRISM/CLEAR ($R = 100$), \nii$\lambda$6584 is blended with $\rm H{\alpha}$. %We first measure the line fluxes of the combined emission from these two lines. %Assuming a dust-free environment where the line ratio \ha/\hb = 2.86, we compute the line flux of \ha. 
We use $A_V=0.30$ from the \bagpipes\ fit of G1 and the derived line ratio \ha/\hb $\sim 3.16$, assuming Case B recombination. We then us this information to compute the line flux of intrinsic $\rm H{\alpha}$.
The difference between the blended line flux and the intrinsic \ha flux then gives us the line flux of \nii$\lambda$6584. Besides \nii$\lambda$6584, other lines of G1 can be measured clearly. 

G1 exhibits high \oiii$\lambda$5007/\hb and low \nii$\lambda$6584/\ha ratios, consistent with findings reported for AGNs at $z=4-7$ \citep{Nakajima2023,Sanders2024,Harikane2023,Kocevski2023}. The AGN-hosting galaxies are hard to distinguish from normal star-forming galaxies at $z > 4$ in the BPT diagram, possibly due to the low metallicities, with $Z/Z_{\odot} \sim 0.19$ for G1 and $0.25$ for G2 from \bagpipes\, fitting, observed in these sources.

In Figure~\ref{fig: BPT}, we use the BPT diagnostic diagram \citep{1987ApJS...63..295V, 2001ApJ...556..121K, 2003MNRAS.346.1055K, 2006MNRAS.372..961K} to determine the presence of AGN in our system. % 
Although it has been shown that the BPT diagram can miss true AGN due to the lack of heavy elements at high-$z$\citep{Curti2023,Kocevski2023}, we use it as another piece of possible evidence for the AGN nature of our systems.
Figure~\ref{fig: BPT} (left panel) is the \oiii$\lambda$5007/\hb vs. \nii$\lambda$6584/\ha diagram for G1, accompanied by two curves at low redshift that differentiate between sources dominantly influenced by AGNs and star formation \citep{Kewley2001,Kauffmann2003}.  From this BPT diagram there is a strong suggestion that G1 likely has an AGN contribution. We can infer that it is a dust-obscured Type-II AGN, with spectral lines produced by highly ionized gas powered by the supermassive black hole at its center. Our AGNs exhibit high \oiii$\lambda$5007/\hb and low \nii$\lambda$6584/\ha ratios, which is consistent with findings reported for AGNs at $z=4-7$ \citep{Nakajima2023,Sanders2024,Harikane2023,Kocevski2023}. 
Figure~\ref{fig: BPT} (right panel) is the \oiii$\lambda$5007/\hb vs. \sii$\lambda$6730/\ha diagram for G1. The \sii line has been rarely observed before at high $z$, but we identify a potential detection with a $S/N\sim 2.8$ in the spectrum from the DAWN JWST Archive (DJA) \footnote{https://dawn-cph.github.io/dja/index.html}. %It is hard to discern this line in the 2D spectrum, but i
%Its peak aligns closely with that of the \ha line. 
%From this BPT diagram w
We find that G1 closely resembles a Seyfert galaxy, where AGN activity contributes partly to the total measured SED and line fluxes but does not dominate over the contributions from star formation.

Additionally, an indicator of AGN activity at high redshifts is the presence of high-ionization lines, such as \NeV,$\lambda$3427 and \HeI,$\lambda$2945, which are typically associated with type-II AGNs \citep{Brinchmann2023, Chisholm2024, Scholtz2023} and signify an environment shaped by intense, energetic processes. %After subtracting the continuum, we integrated over the spectral axis to measure the line fluxes.
We detect \NeV and \HeI in G1. The flux of \NeV is $6.9\pm2.3 \times 10 ^{-19}$ erg s$^{-1}$ cm$^{-2}$ with SNR = 3.0, and \HeI is $9.1\pm2.9 \times 10 ^{-19}$ erg s$^{-1}$ cm$^{-2}$ with SNR = 3.1. %\NeIV has a SNR = 1.8 weak signal with a flux of $4.6\pm2.5 \times 10 ^{-19}$ erg s$^{-1}$ cm$^{-2}$. 
The high line ratio log10(\NeV/\ciii)>0 classifies these objects as AGN host galaxies \citep{Scholtz2023}. %The FRESCO NIRSpec spectrum wavelength range of G2 does not include any high ionization UV lines.

\subsection{\texorpdfstring{Broad H$\alpha$ line and black hole mass}{Broad Halpha line and black hole mass}}

\begin{figure}
    \centering
    \includegraphics[width=0.98\linewidth]{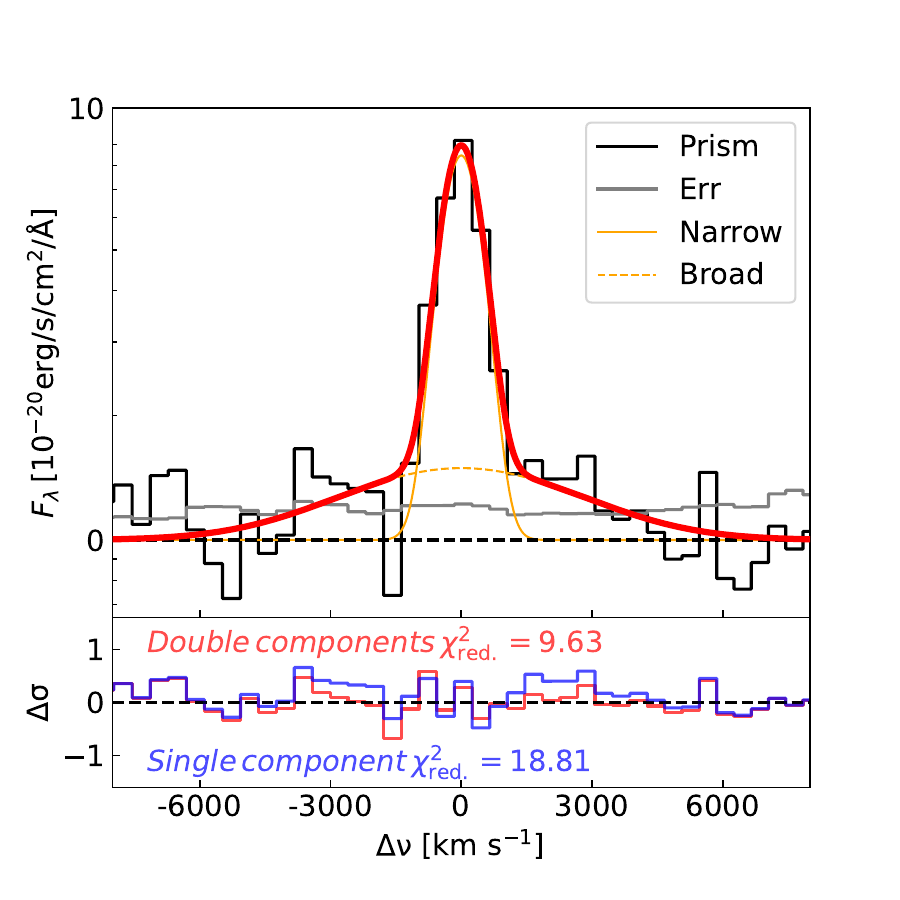}
    \caption{
    The NIRSpec PRISM/CLEAR spectra of G1 are shown as black histograms, with the associated uncertainties shown as gray error bars. We use a double Gaussian model to fit H$\alpha$, shown in the top panel. The best-fit narrow and broad Gaussian components for the \ha emission are represented by the solid and dashed orange lines, respectively. The red line indicates the combined fit of these two components. The bottom panel shows the residuals from the double and single Gaussian fits, with their reduced $\chi_{\rm red.}^2$.    
    }
    \label{fig: two gaussian Ha}
\end{figure}

More evidence supporting the presence of an AGN in G1 is the detection of a broad H$\alpha$ line component. 
The grating R1000 observations are too shallow to detect these lines compared to the lower resolution PRISM data. Therefore, we use the PRISM spectrum of G1 to measure line widths. To fit our spectra and measure \ha line properties, we used both single Gaussian and double Gaussian models with a Markov chain Monte Carlo (MCMC)-based technique, where the line centers, widths, and fluxes are free parameters. The double Gaussian models contain \ha emission from the broad-line region, which is blended with narrow \ha lines from the galaxy. The MCMC process was repeated 10,000 times until convergence was achieved. We also calculated the corresponding $\chi_{\rm red.}^2$ values for the fitting (Figure~\ref{fig: two gaussian Ha}). We note that due to the low resolution, we did not include an \nii $\lambda\lambda$6548, 6583 line in our fit.

In the \ha line profile of G1, for the single Gaussian fit, the best-fit \ha component has a line flux $f_{\rm single} = 2.1\pm0.1\times 10^{-18}\mflux$ SNR = 23.5. $L_{\rm Ha}=7.0\pm0.4\times 10^{41}$erg s$^{-1}$, 
and FWHM = $1141.5^{+68.9}_{-49.0}$ \kms, 
which show a potential broad line component. We then carry out a double Gaussian fit with both narrow and broad Gaussian components. 
For the double Gaussian fit, the narrow component has $f_{\rm narrow} = 1.7\pm0.1\times 10^{-18}\mflux$ SNR = 15.5 and FWHM = $1009^{+34}_{-18}$ \kms, while the broad component has $f_{\rm broad} = 0.8\pm0.2\times 10^{-18}\mflux$ SNR = 3.4 and FWHM = $5804^{+1811}_{-2165}$ \kms.
% calculation results:
%1009.0483954429554 +- [-18.9186596   33.70042851]
%5803.780018514578 +- [-2165.26133605  1811.63311568]
%Line flux for narrow component: 173.27+-11.18 SNR = 15.50
%Line flux for broad component: 75.28+-22.27 SNR = 3.38
The derived broad line luminosity is $L_{\rm Ha,\ broad}=2.6\pm0.8\times 10^{41}$erg s$^{-1}$. The presence of a broad line component with a width of $>5000$ km/s indicates that G1 harbors an AGN. %, with spectral lines produced by highly ionized gas powered by the supermassive black hole at its center, consistent with galaxies or AGNs at $z=4-7$ ({\color{blue} Nakajima et al. 2023; Sanders et al. 2024; Harikane et al. 2023; Kocevski et al. 2023}).

% single:
%1141.4625870197879 +- [-46.47592724  69.87812953]
%Line flux for single component: 205.17+-8.72 SNR = 23.53

% 2306.05448v3.pdf s4.2
Given the broad component is assumed attributable to AGN activity, we can calculate black hole mass using virial relationships and parameters fitted to the broad component \citep{Matthee2024,Harikane2023,Kocevski2023}. We estimate the black hole mass using the equation \citep{Reines2013,Reines2015}: 

\begin{equation}
\begin{split}
\log_{10}\left(\frac{M_{\rm BH}}{M_{\odot}}\right) =\ & 6.57 + \log_{10}(\epsilon) \\
& + 0.47 \log_{10}\left(\frac{L_{H\alpha,\text{broad}}}{10^{42}\, \text{erg s}^{-1}}\right) \\
& + 2.06 \log_{10}\left(\frac{v_{\text{FWHM,broad}}}{10^3\, \text{km s}^{-1}}\right)
\end{split}
\label{eq:black_hole}
\end{equation}

Where we assume $\epsilon = 1$, represents a geometric correction factor for the broad line region. Additionally, the systematic uncertainty estimates using this virial approach is noted to be approximately 0.5 dex \citep{Reines2015}. %The black hole mass estimated from the broad \ha line is $M_{\rm BH} \sim 4.3 \times 10^6 M_{\odot}$.
% log M_star = 7.253 ~ 1.8e7
The black hole mass, estimated from the broad H$\alpha$ line, is $M_{\rm BH} \sim 4.3 \times 10^6 M_{\odot}$, corresponding to approximately 24\% of the host galaxy's stellar mass ($M_{\star} = 1.8 \times 10^7 M_{\odot}$). 
It is comparable to the typical black hole mass derived within the `Little Red Dots' (LRDs) faint AGN at $z \sim 5$ and faint broad-line AGNs at $z=4-7$, which is $M_{\rm BH} = 0.2-9.8 \times 10^7 M_{\odot}$ \citep{Matthee2024,Harikane2023}.

The derived $M_{\rm BH} / M_{\star}$ ratio is 0.24. It is higher than the observed `Little Red Dots'(LRDs) and AGNs at $z \sim 6$ ($M_{\rm BH} / M_{\star} \sim 0.01-0.1$)\citep{Harikane2023,Maiolino2023BL,Andika2024}, suggesting that the black hole is relatively overmassive compared to its host galaxy. This may point to accelerated black hole growth in the galaxies in pairs or an alternative evolutionary pathway, possibly driven by galaxy mergers.

%[NeIV]λ2424 4.629286551761109e-19 1.8330120784620894
%[NeV]λ3427  6.913226688223706e-19 2.961306337010506
%HeIλ2945 9.052244976924936e-19 3.074510417

\subsection{\texorpdfstring{The associated strong extended \lya emission}{The associated strong extended Ly-alpha emission}}

We discover that this dual AGN is associated with strong extended \lya emission from VLT/MUSE observations.  Fig.~\ref{fig: image} b shows the continuum-subtracted, pseudo-narrow band image, spanning $\approx \pm 2000$ km s$^{-1}$ with central wavelength $\lambda_{\rm NB} = \lambda_{\rm Ly\alpha} (1+z_{\rm spec}) \approx 7831 $\AA, and Gaussian smoothed spatially by $1$ arcsec. %Here the $z_{\rm spec}$ is from the \hb measurement. 
The $2\sigma$ surface brightness limit is $9.1\times 10^{-18}$ erg s$^{-1}$ cm$^{-2}$ arcsec$^{-2}$. 
The aperture ($\sim1.7 \rm \, arcsec^2; 1.4'' \times 1.2''$) matches the seeing of the observations. 
From the narrow band image, we find G2 has strong and broad Ly$\alpha$ emission. The peak surface brightness of G2 is $\approx 3.8 \times 10^{-17} \rm \, erg \, s^{-1} \, $ $\rm cm^{-2} \, arcsec^{-2}$. To a 2$\sigma$ surface brightness contour, this source has a size of $\gtrsim 3.7$ arcsec (i.e.\ $\gtrsim 22$ physical kpc) and has a total Ly$\alpha$ luminosity of {$> 3.3\pm 0.1 \times  10^{43} \, \rm erg \, s^{-1}$}.  This makes this system one of the most extended and bright Ly$\alpha$ nebulae at $z\sim6$ discovered (e.g. HSC sample\citep{haibin2020}, MUSE sample\citep{Drake2019}). 

We extract 1D spectra from the VLT/MUSE final datacube for G1 and G2 using an aperture of 0.32'' $\times$ 0.32''.  
We find that G2 shows broad $>5000\mkms$ \lya\ emission (FWHM$_{\rm G2} =372.88 \pm34.35$\AA). G1 do not exhibit Ly$\alpha$ emission, that is modeled by a single Gaussian with central wavelength $\lambda_{\rm Ly\alpha} = 7874.0 \pm 14.8$\AA\ corresponding to $z = 5.48 \pm 0.01$. This might be due to intrinsic weak emission or absorption. {Its 3$\sigma$ upper limit flux within an aperture of $1''\times 1''$ is $\approx 4.9 \times 10^{-18} \rm \, erg \, s^{-1} \, $ $\rm cm^{-2}$.}

The source with strong \lya (G2) shows weak H$\alpha$, while the source without any \lya (G1) exhibits strong \ha, \hb, and \oiii emissions, suggesting G2 has significant dust obscuration along our line of sight. All of these results suggest that both G1 and G2 are capable of producing sufficient ionizing radiation to contribute to the local ionization field. The higher ionizing photon production efficiency of G1 indicates that its ionizing radiation is more intense relative to its UV continuum, aligning with its stronger optical emission lines. Conversely, G2's weaker H$\alpha$ emission, despite its high Ly$\alpha$ escape fraction, suggests that the observed Ly$\alpha$ emission is likely enhanced by additional processes such as anisotropic AGN-driven photoionization or shocks from AGN outflows \citep{Witten2024}, especially given its location within a dense galaxy environment (see Section~\ref{sec: over}). This scenario is consistent with observations and theoretical predictions that galaxy mergers and AGN activity can elevate ionizing photon production and facilitate Ly$\alpha$ photon escape in the early universe\citep{Witten2024}.

\subsection{Overdensity measurements}\label{sec: over}

To ensure a robust evaluation of the galaxy environment, accounting for variations in the selected surrounding area size, we utilize a k-dimensional tree (KDTree) data structure to facilitate the nearest neighbor method to effectively estimate the local environment of each galaxy \citep{Lopes2016}. For each galaxy, we calculate its projected distance, denoted as $d_n$, to the $n$th nearest neighbouring galaxy, considering a maximum velocity offset $\Delta z\,<\,0.1$. The local galaxy density, $\Sigma_n (z)$ is denoted as: 
\begin{equation}
\Sigma_n (z) = n/\pi d_n^2
\end{equation}
where $\Sigma_n$ is expressed in units of galaxies/Mpc$^2$. The choice of the parameter $n$, representing the rank of the density-defining neighbor, is crucial. It is imperative to ensure that $n$ is not greater than the number of galaxies within the halo, as this could lead to a loss of sensitivity in our environmental analysis. We adopt the \sfive, local galaxy density estimator with $n$ set to 5. This choice is often preferred in the literature, as it tends to be smaller than the number of galaxies typically found in a cluster, ensuring a robust and commonly used estimate\citep{Lopes2016,Qiong2024}. 

\begin{figure*}
\centering
\includegraphics[width=0.95\textwidth]{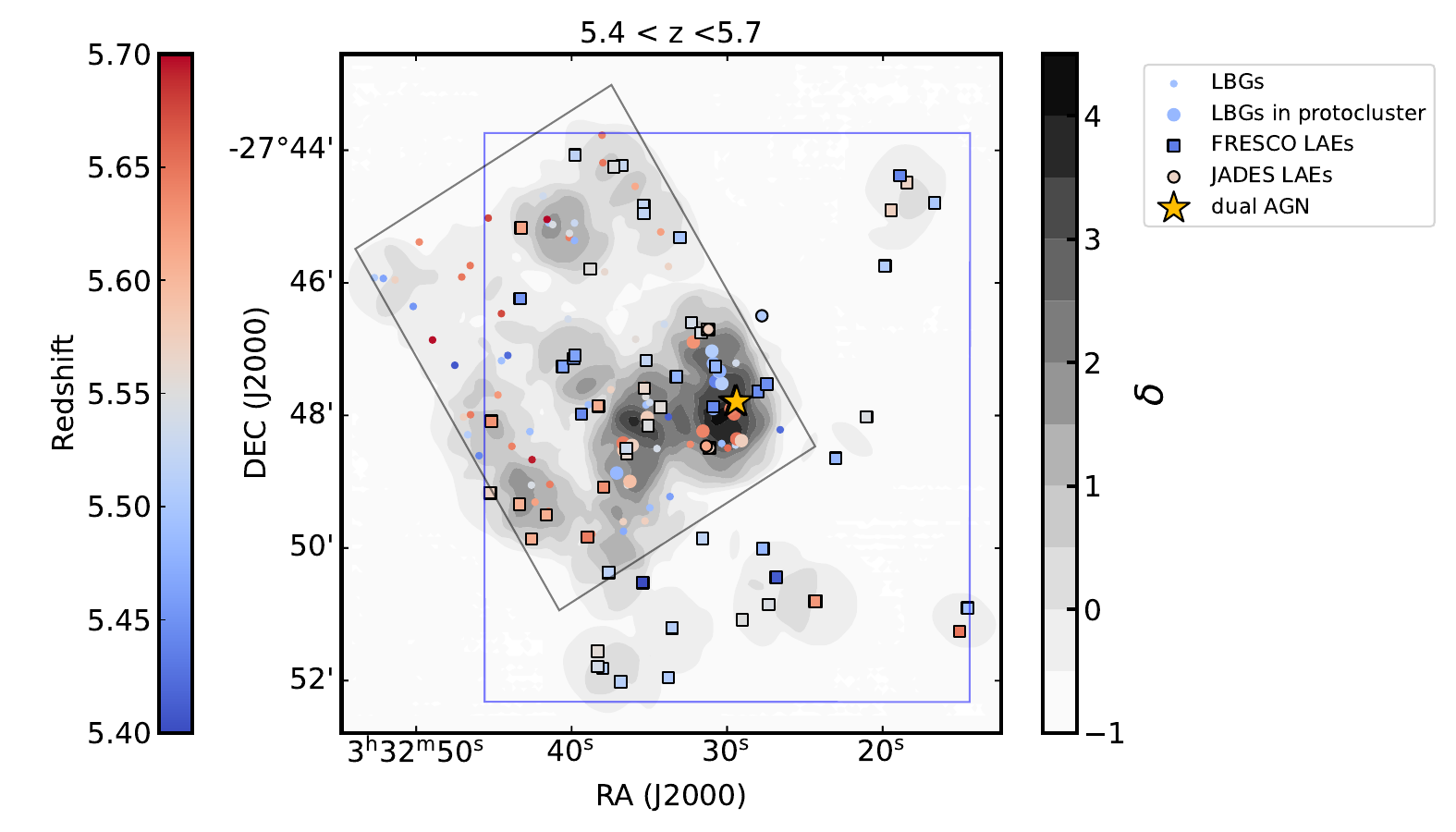}
\caption{
The position of the dual AGN candidate and nearby LAE, LBG systems at $z=5.4-5.7$ overlaid on a galaxy density map. The blue and black rectangles are FRESCO and JADES DR1 footprints. The yellow stars are the dual AGN, which are two close to distinguish; black-bordered circles and squares are spectroscopically confirmed LAEs from the JADES and FRESCO surveys. Circles without black borders are photometric confirmed LBGs. Larger circles indicate members of a protocluster, JADES-ID1-5.68, falling within this redshift range. The overdensity ($\delta$) is represented in grayscale. It is evident that the dual AGNs are located at the highest-density region of this structure, possibly at the node of several filaments. 
}
\label{fig: cluster}
\end{figure*}

When we calculate these overdensites, we find that this dual AGN is associated with a galaxy overdensity\citep{Qiong2024}, potentially located at the center of a $z\sim5.4$ protocluster or filamentary structure node. The 5th nearest neighbor density $\Sigma_5$ are 128 and 110 galaxies/Mpc$^2$ for G1 and G2, in the top 5\% of \sfive distribution. It is 2.2-2.6 times higher than the average \sfive in their redshift range. Figure~\ref{fig: cluster} shows the distribution of dual AGNs associated with a photometric protocluster, along with spectroscopically confirmed LAEs at redshift $z=5.3-5.7$. Spectroscopically confirmed LAEs include those from the JADES and FRESCO surveys, as well as MUSE. 

Our dual AGN is located in the density peak of the photometry-confirmed protocluster, JADES-ID1-5.68, that contains a total of 66 galaxies ($P_{\text{mem}}>0.32$) within the redshift range $5.2 < z_{\text{phot}} < 6.2$. The derived dark matter halo mass of the protocluster is $M_{\rm halo} = 13.28^{+0.37}_{-0.34} M_\odot$ \citep{Qiong2024}, the highest among the JADES fields. It could evolve into clusters with $M_{\rm halo} > 10^{14}M_{\odot}$ at $z = 0$, resembling `Coma' or `Fornax' type clusters \citep{Chiang2013,Qiong2024}. %The dual AGN is also in close proximity to the spectroscopically confirmed protocluster JADES-GS-OD-5.386, which identified 42 spectroscopic sources at redshifts $z=5.2-5.5$\citep{Helton2024}. G2 was included and labeled as `field galaxy'. It could be attributed to limitations in incompleteness sampling of spectroscopic observations.%, and it exhibits a significant deviation from the position of the protocluster JADES-GS-OD-5.386. 

In Figure~\ref{fig: cluster}, we map the overdensity ($\delta$) at $z=5.3-5.7$, defined as $\delta = \frac{N - \overline{N}}{\overline{N}}$, where $N$ and $\overline{N}$ represent the total and average numbers of galaxies in a cylinder, respectively. %The map is generated by smoothing the calculated overdensities with a Gaussian kernel, where the standard deviation is $\sim2$ cMpc at $z\sim5.4$. %We consider the Poisson statistics for $N$ and $\overline{N}$. 
Within a 20 arcsec ($\approx120$~pkpc) region, the overdensity around the dual AGNs is $\delta =4.3$, which is $>5$ times more than the random field. Combining all observations, we clearly find that these dual AGNs appear to be situated at the center of a $z\sim5.4$ protocluster or filamentary structure node.

\section{Discussion}\label{sec: Discussion}

\subsection{Dual AGN and Starburst Nature of the Merging System}
We have shown that SED fitting, BPT diagrams, the broad H$\alpha$ profile, the detection of high-ionization lines, and the remarkably surrounding extended Ly$\alpha$ nebula all provide strong evidence for significant AGN activity in both G1 and G2. To assess the presence of AGN components, we fit the SED using \eazy{} with pure star formation templates, and an AGN+star formation template, including templates for direct collapse black hole (DCBH) hosts from \citet{Nakajima2022} and a set of galaxies templates \citet{larson2022spectral}. The fraction of AGN light from G1 and G2 ($f_{\rm agn}$) in these fits both approach 1 and the $\chi_{\rm red.}^2$ value is notably lower for the AGN than star formation by $\Delta\chi_{\rm red.}^2>8$, strongly suggesting that both sources likely harbor AGNs (Figure~\ref{fig: SED}). 

At the same time, these two galaxies are both starbursting and have low metallicities. We use \bagpipes\, to simultaneously fit photometric and spectroscopic data with a simple AGN model. This AGN model has a broken power-law and broad \ha and \hb components. The burstiness factors (SFR$_{\rm 10 Myr}$/SFR$_{\rm 100 Myr}$) of G1 and G2 are both $\sim$ 7, indicating a strong starburst system. Our fitted metallicities of G1 and G2 are $Z/Z_{\odot} \sim 0.19$ and 0.25, which is consistent with $z>5$ AGN in CEERS \citep{Kocevski2023} and the model predictions for moderately low-metallicity AGN at high redshift with $Z/Z_{\odot} \sim 0.2 - 0.4$. From this {\tt BAGPIPES} SED fitting the SFR of G2 is $\sim$2.1-5.3 M$_{\odot}$ yr$^{-1}$, while G1 is $\sim$1.7-2.6 M$_{\odot}$ yr$^{-1}$. %Their properties are similar to low-z dwarf galaxies. 

The ALMA Science Archive currently contains data covering these objects in ALMA Band 3 and Band 6 with continuum RMSs of 0.015 and 0.035 mJy/beam, respectively. However, no continuum emission was detected. We assume a modified black body dust model with a temperature of 42 K and an emissivity index of 1.6 (typical for AGN-starburst systems\citep{Beelen2006,Leipski2013}), we estimate the limiting FIR luminosity to be $L_{\rm FIR} = 1.0 \times 10^{10} L_{\odot}$, corresponding to the upper limit of SFRs of $\sim$ 12 and 84 $M_{\odot}$yr$^{-1}$, respectively. The absence of detections could be due to the system's low metallicity.

\subsection{The Escape Fraction of Ly\texorpdfstring{$\alpha$}{Lg} photons and Production Efficiency of Ionizing Photons}

We estimate the escape fraction of Ly$\alpha$ photons ($f^{\rm Ly\alpha}_{esc}$) for G2 assuming Case B recombination. We adopt an electron temperature of $T_e = 10^4$ K and an electron density of $n_e = 100 \, \text{cm}^{-3}$, following standard assumptions for ionized nebulae \citep{Osterbrock2006}. 
We apply the standard ratio L(\lya)/L(\ha) = 8.7 (e.g. \citep{Henry2015}) to calculate the intrinsic \lya flux from the observed \ha flux. The estimated $f^{\rm Ly\alpha}_{\rm esc}$ is as: 
\begin{equation}
f_{\rm esc}^{\rm Ly\alpha } = \frac{L_{\rm obs}(\rm Ly\alpha)}{8.7 \times L({\rm H{\alpha}})}
\end{equation}
The 3$\sigma$ upper limit \ha luminosity of G2 is $2.0\times10^{41}$erg $s^{-1}$; with the same square aperture ($0.31''\times 0.31''$) where we extract the H$\alpha$ luminosity, we measured the Ly$\alpha$ luminosity to be $L_{\rm obs}(\rm Ly\alpha)=1.8\pm 0.3 \times 10^{42}$ erg s$^{-1}$.
%$f < 5.9\times10^{-19} \mflux$. 
%We apply a dust correction using the reddening law described by \citet{Calzetti2000}. Given the absence of Balmer decrement ({\ha/\hb}) information to constrain the extinction ($E(B - V)$) for nebulae, we use AV from BAGPIPES SED fitting \citet{Carnall2018}.
% 2.5e44/(2e41*8.7)
The $f^{\rm Ly\alpha}_{esc}$ of G2 is thus $\approx 1$. This suggests that G2 efficiently emits Ly$\alpha$ photons, with minimal attenuation by dust or neutral hydrogen. Such a high escape fraction is rare, particularly at high redshift, and may indicate that G2 has a relatively low neutral hydrogen column density or that the Ly$\alpha$ photons are escaping through channels cleared by AGN-driven outflows. The combination of strong Ly$\alpha$ emission, a high escape fraction, and the presence of an AGN suggests that AGN activity could be playing a role in facilitating Ly$\alpha$ photon escape by altering the local ISM conditions or creating ionized channels through which Ly$\alpha$ photons can escape more freely.

We also estimate the production efficiency of ionizing photons ($\xi _{\rm ion}$), as:
\begin{equation}
\xi _{\rm ion} = \frac{\dot{n}_{\rm ion}}{L_{\nu }^{\rm UV}}
\end{equation}

where \( \dot{n}_{\text{ion}} \) (s\(^{-1}\)) is the intrinsic production rate of hydrogen-ionizing photons originating from stellar populations, while \( L_{\text{UV}} \) (erg s\(^{-1}\) Hz\(^{-1}\)) is the (monochromatic) UV-continuum luminosity per photon frequency derived from the \bagpipes\ fitting. \( N_{\text{ion}} \) can be calculated using H\(\alpha\) emission data assuming Case B recombination \citet{Leitherer1995,Madau1998}, as:

\begin{equation}
\dot{n}_{\rm ion} = \frac{L(\rm H\alpha)}{1-f_{\text{esc}}^{\text{LyC}}} \times 7.35 \times 10^{11} {\rm erg}^{-1}
\end{equation}
The derived production efficiency of ionizing photons \( \xi_{\text{ion},0} \) is estimated as \( \log_{10} \xi_{\text{ion},0} = 26.9 \) for G1 and \( <26.3 \) for G2, under the assumption of \( f_{\text{esc}}^{\text{LyC}} = 0 \). It is consistent with \lya emitters at high-$z$ with \( \log_{10} (\xi_{\text{ion},0} / {\rm Hz\, erg^{-1}}) \sim 25.0 - 26.5\)\citep{Saxena2024}. While this assumption is commonly adopted for stellar sources to infer \(\xi_{\text{ion}}\) from recombination lines (e.g., H$\beta$ or H$\alpha$), it may not strictly hold for AGN, which can have significant LyC escape fractions due to their harder spectra and lower surrounding gas covering fractions. Therefore, these values represent lower limits on the true \(\xi_{\text{ion}}\), and should be interpreted with caution when comparing to normal galsxies. Despite this, the high efficiency in ionizing photon production suggests that this dual AGN could play a significant role in reionizing its surrounding environment.

We did not detect strong \lya emission in G1, which is not uncommon in high-redshift galaxies. The strength of \lya emission in high-redshift galaxies is influenced by factors such as absorption by neutral hydrogen (HI) in the interstellar medium (ISM), properties of the circumgalactic medium (CGM) surrounding galaxies, and orientation or line of sight effects. At high redshifts, the Universe was more neutral, with a higher density of HI in the ISM and IGM along the line of sight. \lya photons emitted by stars can be absorbed by these neutral hydrogen atoms, significantly reducing the observed \lya emission. Another factor is the lack of gas in the CGM. High redshift galaxies might have a less enriched or less dense CGM compared to galaxies at lower redshifts. The presence of a substantial and metal-enriched CGM can facilitate the escape of Lyman-alpha photons by reducing the opacity to these photons. Therefore, the observed Lyman-alpha emission can be weaker or absent. Additionally, the orientation of galaxies relative to the observer can affect observed \lya emission; if the line of sight does not intersect regions of significant \lya emission or if the galaxy is viewed edge-on, the observed \lya emission can be faint or absent.

\subsection{Rarity of Dual AGNs at High Redshift}
The discovery of a dual AGN at $z\sim5.4$ using JWST/NIRSpec offers pivotal insights into the early universe's galaxy and SMBH co-evolution as well as their local environment and intergalactic medium. This system, featuring dual AGN activity and an extended Ly$\alpha$ nebula, resides at the centre of a galaxy group at $z>5$. The detection of dual AGNs at such an early epoch is unexpected, as current cosmological models suggest that the formation of massive black holes and their subsequent mergers should be rare in the first billion years after the Big Bang\citep{Puerto2025,Singh2023}. . The fact that these dual AGNs are already active and situated within a galaxy group suggests that galaxy interactions and mergers, which are typically linked to triggering AGN activity, may have occurred more frequently or earlier than previously thought. 

To estimate the rarity of dual AGN in very low-mass hosts ($M_\star\simeq1.8\times10^{7}\,M_\odot$) at $z\sim5.5$, 
we first take the number density of galaxies in the mass range $M_\star\simeq10^{7}$--$10^{7.5}\,M_\odot$ from the galaxy stellar mass function (GSMF) at $z\sim5$--6 \citep[e.g.][]{Weibel2024,harvey2024epochs}. 
We then multiply by the probability that such a galaxy has a physical companion within $r_{\rm max}=10$~pkpc, 
by the massive black hole (MBH) occupation fraction in dwarfs, 
by the AGN duty cycle for such MBHs (squared to require both nuclei to be active), 
and by the fractional lifetime of the kpc-scale dual phase relative to the Hubble time at $z\sim5.5$. The expression is
\begin{equation}
\begin{split}
n_{\rm dual\,AGN}(M_\star,z) \approx 
\phi_\star(M_\star,z) \,\Delta\log M_\star \times 
f_{\rm pair}(<r_{\rm max}) \\ \times 
f_{\rm occ}^2 \times 
f_{\rm duty}^2 \times 
\frac{t_{\rm kpc}}{t_{\rm H}(z)} ,
\end{split}
\label{eq:mass_conditioned}
\end{equation}
where $\phi_\star(M_\star,z)$ is the GSMF, 
$\Delta\log M_\star$ is the stellar mass bin width, 
$f_{\rm pair}(<r_{\rm max})$ is the fraction of galaxies with a 3D companion within $10$~pkpc \citep[e.g.][]{Duan2025}, 
$f_{\rm occ}$ is the MBH occupation fraction \citep[e.g.][]{Bellovary2019}, 
$f_{\rm duty}$ is the single-nucleus AGN duty cycle \citep[e.g.][]{Reines2013}, 
$t_{\rm kpc}$ is the kpc-scale dual phase duration, 
and $t_{\rm H}(z)$ is the Hubble time \citep[e.g.][]{Capelo2017,Steinborn2016}.

For the faint-end GSMF at $z\sim5$--6, JWST measurements give $\phi_\star\simeq(3\times10^{-3}$--$1\times10^{-2})\,{\rm Mpc}^{-3}\,{\rm dex}^{-1}$ in this mass range \citep{Weibel2024}, and we adopt $\Delta\log M_\star\simeq0.5$.  
The close-pair fraction for such low mass galaxies at sub–10~pkpc separations is not directly measured; guided by JWST close-pair studies and small-scale clustering scalings, we adopt $f_{\rm pair}\sim0.005$--0.05 \citep{Duan2025,Eftekharzadeh2017}.  
Simulations and observations suggest $f_{\rm occ}\sim0.3$--1.0 \citep{Bellovary2019}, and for the duty cycle we use $f_{\rm duty}\sim0.05$--0.2 based on dwarf AGN fractions \citep{Reines2013}.  
Hydrodynamical merger simulations indicate $t_{\rm kpc}\sim30$--$100$\,Myr, corresponding to $t_{\rm kpc}/t_{\rm H}(z\!\sim\!5.5)\sim0.03$--0.1 \citep{Capelo2017,Steinborn2016}. It gives 
$
n_{\rm dual\,AGN} \sim (1\times10^{-10} \text{--} 3\times10^{-8})\ {\rm Mpc}^{-3},
$
with the lower (upper) bound reflecting the most conservative (optimistic) combination of parameters.  
For the JADES-like volume $V\simeq2.7\times10^{4}\,{\rm Mpc}^3$, 
$
N_{\rm exp} \sim (3\times10^{-6} \text{--} 8\times10^{-4}),
$
which implies a Poisson probability $P(\ge1)\simeq N_{\rm exp} \ll 10^{-3}$ across conservative assumptions.  
This confirms that the detection of a dual AGN in such low-mass hosts at $z\sim5.5$ is intrinsically very rare.

\section{Conclusions}\label{sec: conclusion}

In summary, we present the discovery of a spectroscopically confirmed dual AGN at $z\sim5.4$ with a projected separation of $\sim10.4$~pkpc, among the highest-redshift examples known. The system is embedded in a prominent overdensity, likely marking the core of a protocluster or a node of a filamentary structure. Both nuclei show clear AGN signatures from SED fitting, emission-line diagnostics, and the detection of high-ionization lines. VLT/MUSE data reveal an extended Ly$\alpha$ nebula exceeding $22$~kpc, consistent with AGN-driven photoionization or shocks, and indicative of substantial circumgalactic gas reservoirs.

The simultaneous activity of two SMBHs in such low-mass hosts, combined with high black hole-to-stellar mass ratios, high Ly$\alpha$ escape fraction, and elevated ionizing photon production efficiency, provides stringent constraints on models of black hole fueling, AGN feedback, and early SMBH growth. The rarity of such systems in current cosmological simulations suggests that galaxy interactions and mergers triggering AGN activity may have been more common and occurred earlier than previously assumed, with dense environments accelerating SMBH mergers and galaxy assembly.

Further deep JWST/NIRSpec spectroscopy will be essential to refine the broad H$\alpha$ measurement, investigate the \oiii+\hb complex and other high-ionization lines in G2, and better constrain AGN properties. Future expanding this search to the remaining unconfirmed AGN–galaxy pairs at $z>4.5$ will enable a statistical assessment of the frequency and impact of dual AGN on galaxy and large-scale structure evolution during the first billion years of cosmic history.

\section*{Acknowledgements}

We acknowledge support from the ERC Advanced Investigator Grant EPOCHS (788113) and support from STFC studentships. Y. Ning acknowledges support from the National Science Foundation of China (no. 12403014). This work is based on observations made with the NASA/ESA \textit{Hubble Space Telescope} (HST) and NASA/ESA/CSA \textit{James Webb Space Telescope} (JWST) obtained from the \texttt{Mikulski Archive for Space Telescopes} (\texttt{MAST}) at the \textit{Space Telescope Science Institute} (STScI), which is operated by the Association of Universities for Research in Astronomy, Inc., under NASA contract NAS 5-03127 for JWST, and NAS 5–26555 for HST. The observations used in this work are associated with JWST program 1345, 1180 1176, and 2738. The authors thank all involved in the construction and operations of the telescope as well as those who designed and executed these observations. 

%%%%%%%%%%%%%%%%%%%%%%%%%%%%%%%%%%%%%%%%%%%%%%%%%%
\section*{Data Availability}

The JWST data used in this work are available in The \textit{JWST} Advanced Deep Extragalactic Survey \citep[JADES;][]{Rieke2023, bunker2023jades, DEugenio2024} (PI: Eisenstein, N. Lützgendorf, ID:1180, 1210). HST data is from Hubble Legacy Fields team \citep{Illingworth2013,Whitaker2019}, which be shared on reasonable request to that team. ALMA data covering these objects in two frequency bands are available from the ALMA Science Archive. The Band 3 data (84.2–115.0~GHz) were obtained under programs 2017.1.00138.S and 2021.1.00547.S, while the Band 6 data (244.1–275.0~GHz) come from programs 2015.1.00098.S, 2015.1.00543.S, and 2017.1.00755.S. In addition, the two sources discussed in this study are covered by the MUSE-Wide guaranteed time observations (GTO) survey with the ESO-VLT \citep{Herenz2019}. Additional data products generated in this work will be made available by the first author upon reasonable request.



\bibliographystyle{mnras}
\bibliography{example} % if your bibtex file is called example.bib

%\begin{thebibliography}{99}
%\end{thebibliography}

%%%%%%%%%%%%%%%%%%%%%%%%%%%%%%%%%%%%%%%%%%%%%%%%%%

%%%%%%%%%%%%%%%%% APPENDICES %%%%%%%%%%%%%%%%%%%%%

%\appendix


%If you want to present additional material which would interrupt the flow of the main paper,it can be placed in an Appendix which appears after the list of references.

%%%%%%%%%%%%%%%%%%%%%%%%%%%%%%%%%%%%%%%%%%%%%%%%%%

% Don't change these lines
\bsp	% typesetting comment
\label{lastpage}
\end{document}